\documentclass{revtex4}
\usepackage{graphics}
\usepackage{epsfig}
\begin{document}

\newcommand{\be}{\begin{equation}}
\newcommand{\ee}{\end{equation}}
\newcommand{\nn}{\nonumber}
\newcommand{\bea}{\begin{eqnarray}}
\newcommand{\eea}{\end{eqnarray}}

\title{Matter effects and CP violating neutrino oscillations \\ with non-decoupling heavy neutrinos}
\author{B.~Bekman, J.~Gluza, J.~Holeczek, J.~Syska, M.~Zra\l ek}

\affiliation{Institute  of Physics, University of
Silesia, Uniwersytecka 4, PL-40-007 Katowice, Poland }

\date{\today}

\begin{abstract}\noindent
The evolution equation for active and  sterile  neutrinos propagating in general
anisotropic or polarized background environment is found and solved for  a special case
when  heavy neutrinos do not decouple, resulting in  non-unitary mixing among light neutrino states. 
Then new CP violating neutrino oscillation  effects appear.
In contrast to the standard unitary neutrino oscillations these effects can be visible
even for two flavour neutrino transitions and even if one of the elements of the neutrino mixing matrix is equal to zero.
They do not necessarily vanish with $\delta m^{2} \rightarrow 0$ and they are  different for various pairs of flavour
neutrino transitions ($\nu_e \to \nu_\mu$), ($\nu_\mu \to \nu_\tau$), ($\nu_\tau \to \nu_e$).
Neutrino oscillations in vacuum and Earth's matter are calculated  for some fixed baseline 
experiments and a comparison between 
unitary and non-unitary oscillations are presented. It is shown, 
taking into account the present experimental constraints, that 
heavy neutrino states can affect CP and T asymmetries. This is especially true in the case
of $\nu_\mu \to \nu_\tau$ oscillations.
\end{abstract}


\maketitle


\section{Introduction}
In the last few years a common effort of experimentalists and
theoreticians yield new informations on the neutrino sector. Nowadays we
can say that  masses and mixing of the three light neutrinos are
quite well established. The unknown issue is  the total
number of neutrino species \cite{est}. Moreover,  additional neutrinos can
be both light (with masses of the order of electronovolts) or heavy (with
masses greater than  the Z boson mass)
\cite{review}. In both cases their couplings to the ordinary
matter must be much smaller than the couplings of the three known
neutrinos. Additional light neutrinos are called
sterile. Such neutrinos are still permissible  from the point of view of primordial nucleosynthesis \cite{pri} and
seem to be necessary if LSND results \cite{lsnd} 
are confirmed by MiniBOONE \cite{miniB}. 
As far as
heavy neutrinos are concerned, their number is not established.
For cosmological reasons they must be unstable \cite{cos}  or they
do not exist at all besides the heavy compact objects with the most 
exotic physics \cite{jacek}.

Let $U_\nu$ be the full neutrino mixing matrix and let the submatrix ${\cal
U}$ (of dimensions $(3 + n_{s}) \times (3 + n_{s})$) constitute the mixing matrix of
light known neutrino states and let the submatrix ${\cal  V}$ (of dimensions $(3 + n_s) \times n_R $)
be responsible for the mixing of the light neutrinos with $n_R$
additional heavy states. Then

\begin{equation}
\label{U-first} U_\nu = \left( \matrix{ {\cal U} & {\cal V} \cr
                        {\cal V}' & {\cal U}'} \right).
\label{gener}
\end{equation}

Many new physics models predict heavy neutrinos. In case they are 
very heavy objects at the unification scale, the
see-saw mechanism explains why the known neutrinos are light and
the matrix ${\cal  U}$, which enters neutrino oscillation
formula, is practically unitary (light-heavy neutrino mixing
representing by the ${\cal  V}$ submatrix is negligible). Without the
see-saw mechanism the submatrix ${\cal  V}$ could have substantial
elements. This situation might be realized with TeV neutrino
masses. However, even then the construction of the matrix $U_\nu$ is not trivial
\cite{tev}, unless all elements of the neutrino mass matrix are of the same order of magnitude and symmetries
disconnect light-heavy mixing from the ratio of their masses \cite{lang}.

Nowadays, experimental data constraints 
elements of ${\cal  V}$ by terms of the following form \cite{lang}
\begin{equation}
\label{c-first} c_{\alpha \beta} \equiv \left( {\cal  V}{\cal
V}^\dagger \right)_{ \alpha \beta}=\sum\limits_{i=heavy} {\cal
V}_{\alpha i} {\cal  V}_{\beta i}^{\ast},
\end{equation}
so the submatrix ${\cal  U}$, which decides about the light neutrino
oscillation effects is not unitary, and nonunitarity effects are
conjugate with $c_{\alpha \beta}$
\begin{equation}
   \left( {\cal U}{\cal U}^\dagger \right)_{ \alpha \beta}=\delta_{ \alpha \beta}-c_{ \alpha \beta}.
\end{equation}

This means that various channels of neutrino oscillations will
depend on the $c_{\alpha \beta}$ nonunitarity  parameters. Some  of such
studies have already been done \cite{lang,val}. Here we would like to concentrate on the influence of the
non unitary matrix ${\cal  U}$ on CP-violating neutrino
oscillations. In the previous paper \cite{acta} we have found
that CP violating oscillations in vacuum can be affected by
$c_{\alpha \beta}$ parameters. We know that, for conventional
unitary neutrino oscillations, the CP violation  can occur
for three (or larger) number of mixed neutrinos, and is small
(vanishes)  if any single element of the ${\cal  U}$ matrix is
small (vanishes). The CP violating effects are govern by one (three) Jarlskog
invariant(s) \cite{jar} for three (four) neutrinos. Finally, they vanish for
short baseline oscillations \cite{cprev}. All such limitations
can be avoided in the case of nonunitary neutrino oscillations. 

In
reality, the influence of matter effects on neutrino
oscillations is important. It can happen that the matter
effects  mimic or even screen  CP effects \cite{cpmat}.
We will also investigate the role of the non-unitary ${\cal
U}$ matrix in the neutrino  oscillations in matter.

The paper is organized as follows. First, in Chapter II the
equation of motion for neutrino states in medium is derived. Such
an equation is well known in the case of the unitary neutrino mixing. To
the best of our knowledge the same equation has never really been found for the
non-unitary mixing. Next, we use this equation to find
various CP and T asymmetries, at first analytically (Chapter~III)
and then numerically (Chapter~IV). Conclusions close the paper.

\section{Propagation of states in matter}

It is well known that
the interference between scattered and unscattered neutrinos can
be crucial for their propagation in matter, even if the probability of
incoherent  neutrino scattering is negligibly small. There are several
derivations of the evolution equation for neutrinos in matter
\cite{sev}.
In all cases, at first an effective potential $V_{eff}$, which
describes the averaged, coherent neutrino interactions with all
background particles, has to be calculated. Using this potential,
the Dirac's equation for a neutrino bispinor wave function $\Psi$
can be written
\begin{equation}
\left( i \gamma^\mu \partial_\mu - m - V_{eff} \right) \Psi =0.
\end{equation}

Taking into account that during the evolution of the flavour
states, particle-antiparticle mixing is negligible, neutrinos do not
change their spin projection and that they are relativistic particles,
a simpler, Schr\"odinger like evolution equation  can be found

\begin{equation}
i \frac{d}{dt} | \nu(t) \rangle = H_{eff}   | \nu(t) \rangle .
\label{sch}
\end{equation}

Usually the effective potential $V_{eff}$ is calculated in the
neutrino flavour basis. This is the traditional approach to the three active
neutrino mixing. If sterile and/or heavy neutrinos exist, it appears to
be more natural to calculate  $V_{eff}$ in the eigenmass basis.
There are several reasons why it is so. At first, it is not
clear conceptually how to define, in a consistent way, creation and annihilation
operators for flavour states \cite{conc}. At second, as a matter of fact,
there is no such an object  as a flavour eigenstate. For quarks,
for instance, only eigenmass states are used. In this context the
only difference between quarks and neutrinos are much smaller
$\delta m^2$'s. However, as we will see later, the most important thing
is that: using the eigenmass basis we will be able to avoid the non-hermitian evolution of neutrino
flavour states affected by the heavy neutrino sector in a matter.
The equation of motion in Chapter III can be  described by a hermitian Hamiltonian from which  real 
effective neutrino masses follow. Final probabilities of flavour changing are affected by the heavy neutrino sector through initial conditions 
and then nonunitary effective neutrino mixing appears.

\begin{figure}[ht]
\epsfig{figure=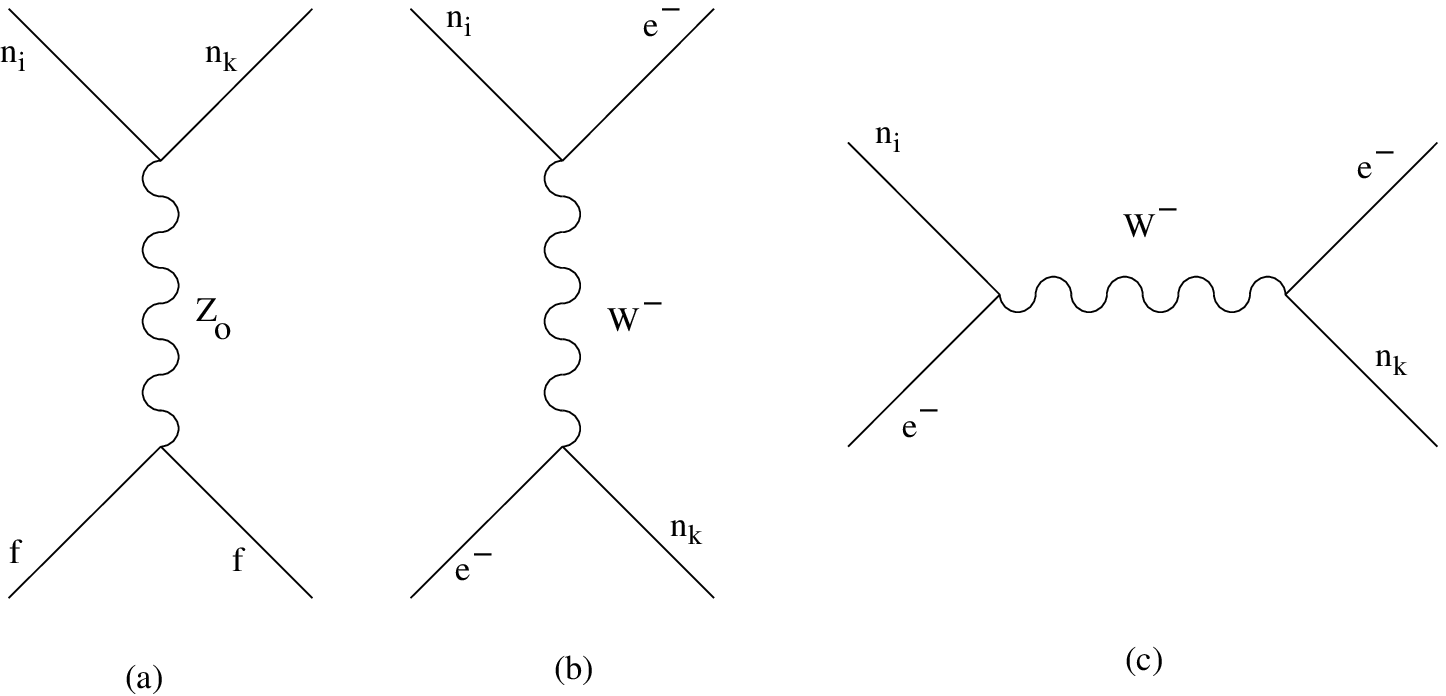, width=10 cm}
\caption{ 
Feynman diagrams for 
scattering of neutrinos in matter . All three diagrams
contribute to neutrino-electron  scattering $n_i+e^- \to n_k+e^-$
$(f=e)$. Only diagram (a) contributes
to  neutrino-nucleon  scattering $n_i+f \to n_k+f$ $(f=p,n)$.} \label{diag}
\end{figure}

In order to find the effective potential  $V_{eff}$, let us assume neutrino interactions in a general form

\be L_{CC}=\frac{e}{2 \sqrt{2} \sin \Theta_W}
\sum\limits_{\alpha=e,\mu,\tau} \sum\limits_{i=1}^n
\bar{\Psi}_{\alpha} \gamma^{\mu} \left( 1 - \gamma_5 \right)
(U_{\nu})_{\alpha i} n_i W_\mu^- + h.c, \label{cc} \ee

and

\be \label{LNC} L_{NC}=\frac{e}{4  \sin \Theta_W \cos \Theta_W}
\left\{ \sum\limits_{i,j=1}^n \bar{n}_{i} \gamma^{\mu} \left( 1 -
\gamma_5 \right) \Omega_{ij} n_j Z_\mu
+ 2 \sum\limits_{f=e,p,n}   \bar{\Psi}_{f} \gamma^{\mu} \left[
T_{3f} \left( 1 - \gamma_5 \right) -2 Q_f \sin^2 \Theta_W
\right]  \Psi_f Z_\mu \right\} \label{nc}  , \ee where $n$ is the
number of light $(3+n_s)$ and heavy $(n_R)$ neutrinos
$(n=3+n_s+n_R)$, and $\Omega_{ij}=\sum_{\alpha=e,\mu,\tau} {(U_\nu)}_{\alpha i}^\ast {(U_\nu)}_{\alpha j}$. 
The coherent neutrino scattering is described in
general by three types of Feynman diagrams presented in
Fig.~\ref{diag}. The Higgs' particles exchange diagrams do not 
contribute as neutrinos are relativistic particles.
In the normal matter all diagrams contribute
to the neutrino-electron scattering $n_i+e^- \to n_k+e^-$ $(f=e)$, yet
only diagram (a) contributes to neutrino-nucleon scattering $n_i+f \to n_k+f$ $(f=p,n)$.

At low energies ($q^{2} \ll M_{W}^{2}, M_{Z}^{2}$), the effective
interaction of light neutrinos 
with a background particle $f$
$(f=e,p,n)$ (Fig.~\ref{diag}), after the appropriate Fierz
rearrangement, can be written in the form

\be H_{int}^f(x)=\frac{G_F}{\sqrt 2} \sum\limits_{i,k=1}^{3+n_{s}}
\sum\limits_{a=V,A} \left[ \bar{n}_k \Gamma_a n_i \right] \left[
\bar{\Psi}_f \Gamma^a \left( g_{fa}^{ki}+{\bar{g}}_{fa}^{ki}
\gamma_5 \right) \Psi_f \right], \ee where
$\Gamma_{V(A)}=\gamma_\mu (\gamma_\mu \gamma_5)$. The couplings
$g_{fa}^{ki}$ and ${\bar{g}}_{fa}^{ki}$ can be calculated from
Eqs.~\ref{cc},\ref{nc} and for electrons $(f=e)$ and nucleons $(f=p,n)$ they are given by

\bea
g_{eV}^{ki} &=&-{\bar{g}}_{eA}^{ki}={\cal U}_{ek}^\ast {\cal
U}_{ei} +
\rho \Omega_{ki} \left( -\frac{1}{2}+ 2 \sin^{2} \Theta_W \right), \nn \\
{\bar{g}}_{eV}^{ki}&=&-g_{eA}^{ki}=-{\cal U}_{ek}^\ast {\cal
U}_{ei}+\frac 12 \rho \Omega_{ki}, \nn \\
g_{fV}^{ki} &=&-{\bar{g}}_{fA}^{ki}=\rho \Omega_{ki} \left( T_{3f} - 2 Q_f \sin^{2}\Theta_W \right), \nn \\
{\bar{g}}_{fV}^{ki}&=& - g_{eA}^{ki}=-\rho \Omega_{ki} T_{3f},
\eea

where

\be
\rho=\frac{M_W^2}{M_Z^2 \cos^2 \Theta_W},T_{3p}=-T_{3n}=1/2,Q_p=1,Q_n=0.
\ee

The global effect of matter - light neutrino interaction can be described by the Hamiltonian

\be H_{int}(x)=\sum\limits_{i,k=1}^{3+n_s} \bar{n}_k V_{ki} n_i \, , \ee

where

\be
V_{ki}=\sum\limits_{f}V_{ki}^f=\sum\limits_{f} \sum\limits_{a} \Gamma^a \left( V_a^f \right)_{ki} \, ,
\label{vfki}
\ee

and

\be
\left( V_a^f \right)_{ki}=\frac{G_F}{\sqrt 2} \sum\limits_{\vec \lambda} \int \frac{d^3p}{\left( 2 \pi \right)^3} \rho_f ( \vec p , \vec \lambda )
 \left( M_a^f \right)_{ki}.
\label{vm} \ee
$ \left( M_a^f \right)_{ki}$ is the  part of the matrix element of the scattering amplitude
$n_i + f \to n_k +f$ connected with the fermion $f$ in the case
 when particles' momenta and spins are untouched

\be \left( M_a^f \right)_{ki} = \langle f, \vec p , \vec \lambda
| \bar{\Psi}_f \Gamma_a \left(  g_{fa}^{ki} + {\bar{g}}_{fa}^{ki}
\gamma_5 \right) \Psi_f | f, \vec p , \vec \lambda \rangle . \ee

In Eq.~\ref{vm}, $ \rho_f ( \vec p , \vec \lambda ) $ is  the
distribution function for  the background particles of spin
$\vec \lambda$ and momentum $\vec p$, normalized in such a way
that $N_f$, defined as

\be N_f \equiv \sum\limits_{\vec \lambda} \int \frac{d^3p}{\left( 2 \pi
\right)^3} \rho_f ( \vec p , \vec \lambda )  
\label{norm} \ee
is the number of fermions $f$ in a unit volume
$(V=1)$. Hence, the amplitude $ \left( M_a^f \right)_{ki}$ must be
calculated for a single fermion in $V=1$ (then for bispinors $u$ we
have $u^\dagger u=1$) and \cite{ber}

\be \left( M_V^f \right)_{ki}^\mu=- \left( M_A^f
\right)_{ki}^\mu= g_{fV}^{ki} \left( \frac{p^\mu}{E_f} \right) +
m_f  {\bar{g}}_{fV}^{ki} \left( \frac{S_f^\mu}{E_f} \right) \, ,
\label{mva} \ee where $E_f,m_f$ and $S_f^\mu = \frac{1}{m_f}
\left( \vec p \vec \lambda, \vec \lambda m_f+ \frac{\vec p ( \vec
p \vec \lambda)}{m_f+E_f} \right)$ denote the  fermion's $f$ energy,
mass and spin four-vector, respectively. The obtained relation
between the vector $M_V$ and the axial-vector $M_A$ amplitudes is the
consequence of the V-A form of interactions
(Eqs.~\ref{cc},\ref{nc}). Now, using Eqs.~\ref{vm} and \ref{mva}
we can write the effective potential $V_{ki}^f$ (Eq.~\ref{vfki})

\bea
V_{ki}^f&=&\left( A_f^\mu \right)_{ki} \gamma_\mu P_L \nn \, , \\
\left( A_f^\mu \right)_{ki} &=& \sqrt 2 G_F N_f \left[
g_{fV}^{ki} \langle \frac{p_f^{\mu} }{E_f} \rangle +  m_f
{\bar{g}}_{fV}^{ki}  \langle \frac{S^\mu_f }{E_f} \rangle \right],
\eea where the average value $<z>$ defined by

\be
<z>=\frac{1}{N_f} \sum\limits_{\vec \lambda} \int  \frac{d^3p}{\left( 2 \pi \right)^3} \rho_f ( \vec p, \vec \lambda )  z ( \vec p, \vec \lambda )
\ee
\\
describes quantities averaged over the fermion distributions $\rho_f ( \vec p, \vec \lambda )$.
In this way the set of coupled Dirac's equations for all light neutrinos propagating in the matter is obtained

\be \sum\limits_{i=1}^{3+n_s} \left( i \gamma^\mu \partial_\mu
\delta_{ki}-m_k \delta_{ki}-V_{ki} \right) n_i =0, \;\;\; k =
1,...,3+n_s. \label{dir} \ee

As we describe the propagation of light relativistic neutrinos only, a 
simpler Schr\"odinger-like evolution equation can be found. Assuming that 
$k^2 \simeq E_i^2 >> m_i^2
\simeq m_i |V_{ki}| \simeq |V_{ki}|^2$, we can get, in the momentum representation,  the Schr\"odinger like equation for each left-handed
(or each right-handed) components of neutrino bispinors 

\be
i \frac{d}{dt} \Psi_k(\vec k, t) = \sum\limits_{i=1}^{3+n_s} H_{ki}^{eff} (t) \Psi_i(\vec k, t).
\label{s1}
\ee

The wave function $\Psi_j(\vec k, t)$ is the neutrino
(antineutrino) state $| \Psi (t)>$ with  momentum $ \vec k$ and
helicity $\lambda =-1 (+1)$, written in the eigenmass basis $|
\nu_j >$, $\Psi_j = <\nu_j | \Psi (t)>$. The effective Hamiltonian (we will assume from now on that all
neutrinos have the same momentum but different energies
$E_j=\sqrt{\vec k^2+m_j^2}$) is equal to 

\be H_{ki}^{eff}=\left( k +
\frac{m_{i}^2}{2k} \right) \delta_{ki}+H_{ki}^{int} \; ,
\label{h1} \ee 

where 

\be H_{ki}^{int} = < \nu_k | \int_{V=1}
d^{3}x \; H_{int}(x)\;  | \nu_i
> = \left\{
\begin{array}{ll}
A^{\mu}_{ki} \bar{u}_k \gamma_\mu P_L u_i &  \mbox{\rm for Dirac neutrinos,} \\
& \\
- ({A^{\mu}_{ki}})^\ast \bar{u}_k \gamma_\mu P_R u_i &  \mbox{\rm for Dirac antineutrinos,} \\
& \\
A^{\mu}_{ki} \bar{u}_k \gamma_\mu P_L u_i-({A^{\mu}_{ki}})^\ast
\bar{u}_k \gamma_\mu P_R u_i &  \mbox{\rm for Majorana neutrinos,}
\end{array}
\right. \ee
and  $A_{ki}^\mu=\sum_f \left( A_f^\mu
\right)_{ki}$.

In the relativistic limit, with the additional assumption
$\vec k \equiv \vec k_i = \vec k_f$, we have

\be {\bar{u}_k \gamma_{\mu} P_L u_i}_{|\lambda=-1} = {\bar{u}_k
\gamma_{\mu} P_R u_i}_{|\lambda=+1}=\left( 1, - \frac{\vec k}{|
\vec k |} \right) \ee
and
\be {\bar{u}_k \gamma_{\mu} P_L
u_i}_{|\lambda=+1} = {\bar{u}_k \gamma_{\mu} P_R
u_i}_{|\lambda=-1}=\left( 0, {\vec 0} \right), \ee
so, finally, we have arrived $(A^\mu_{ki}=(A_{ki}^0,\vec{A}_{ki}))$ to the formula

\be
H_{ki}^{int} = \left\{
\begin{array}{ll}
A_{ki}^0- \frac{\vec k}{| \vec k |} \vec{A}_{ki} &
\mbox{\rm for Dirac and Majorana neutrinos with } \lambda = -1, \\
-({A^{0}_{ki}})^\ast +  \frac{\vec k}{| \vec k |}
(\vec{A}_{ki})^\ast &  \mbox{\rm for Dirac antineutrinos and
Majorana neutrinos with } \lambda = +1.
\end{array}
\right.
\label{effh}
\ee

We can clearly see that the effective interaction Hamiltonian is the same for
Dirac and Majorana neutrinos and that 

\be
H^{int}_{particle} = - [H^{int}_{antiparticle}]^\ast .
\ee

We would like to stress, however,  that these properties are not the general
rules. They are satisfied because of the V-A type of neutrino
interactions in Eqs.~\ref{cc},\ref{nc}, as it is in the case of 
relativistic neutrinos for which the scalar and pseudo-scalar
terms can be neglected.





Eq.~\ref{effh} gives the most general Hamiltonian 
for an arbitrary number of ($3+n_s$) light relativistic neutrinos propagating in any 
background medium and interacting in the $V-A$ way. 
In what follows we
will concentrate on the case of unpolarized
$(<\vec{\lambda}_f>=0)$, isotropic  $(<\vec{p}_f>=0)$ and electrically
neutral ($N_e=N_p \neq N_n$) medium. Then we arrive to the following
Hamiltonian (later we will usually put $\rho=1$)

\be H^{int}_{ki}=\sqrt{2} G_F \left[ N_e {\cal U}_{ek}^\ast {\cal
U}_{ei} - \frac 12 \rho N_n \Omega_{ki} \right] . \label{hint} \ee

The above Hamiltonian always has $(3 + n_{s}) \times (3 + n_{s})$ dimensions,
independently if heavy neutrinos exist or not.
Let us first consider two conventional cases with none and with a
single sterile neutrino, when no heavy neutrino exists ($n_R=0$,
$n=3+n_s$). Then (in both cases)

\be
| \nu_\alpha > =  \sum\limits_{i=1}^n
(U_{\nu}^{\ast})_{\alpha i} | \nu_i >  =
\sum\limits_{i=1}^{3+n_{s}} {\cal U}_{\alpha i}^\ast | \nu_i >
\label{ufbase}
\ee

If no sterile neutrinos are present ($n_s=0$),
then, after removing the redundant diagonal terms in Eqs.~\ref{h1},\ref{hint},
the well known Hamiltonian is obtained ($\alpha, \beta = e , \mu ,
\tau $)

\be H_{\alpha \beta}=\left( {\cal U} \frac{\Delta m^2}{2E} {\cal
U}^\dagger \right)_{\alpha \beta } + \sqrt{2} G_F  N_e
\delta_{\alpha e} \delta_{\beta e}. \label{h2} \ee

If a single sterile neutrino is present ($n_s=1$),
then another well known Hamiltonian is obtained ($\alpha, \beta = e , \mu , \tau , s$)

\be H_{\alpha \beta}=\left( {\cal U} \frac{\Delta m^2}{2E} {\cal
U}^\dagger \right)_{\alpha \beta } + \sqrt{2} G_F  \left[ N_e
\delta_{\alpha e} \delta_{\beta e}+ \frac 12 N_n  \delta_{\alpha
s} \delta_{\beta s} \right]. \label{h3} \ee
\\
From now on, we will only consider cases when there is at least one
non-decoupling heavy neutrino present ($n_R \geq 1$).

If we want to use the full ($3 + n_{s} + n_{R}$) eigenmass basis, we
need to expand the $(3 + n_{s}) \times (3 + n_{s})$ Hamiltonian $H_{ki}^{eff}$, given by
Eqs.~\ref{h1},\ref{hint}, to proper $(3 + n_{s} + n_{R}) \times (3 + n_{s} +
n_{R})$ dimensions, adding zeros

\be
H_{ki}^{eff} \to \left\{
\begin{array}{ll}
H_{ki}^{eff} \mbox{\rm as given by Eqs.~\ref{h1},\ref{hint}} , &
\mbox{\rm if both } k,i \leq (3+n_s) ,\\
0 , &
\mbox{\rm if any of } k,i > (3+n_s) .\\
\end{array}
\right.
\label{fhint}
\ee

These zeros mean that, for the physics of light neutrinos which we are
interested in, the energy and momentum conservation do not allow heavy
neutrinos to be produced nor detected.
If we now define the flavour basis as

\be | \nu_\alpha > =  \sum\limits_{i=1}^n
(U_{\nu}^{\ast})_{\alpha i} | \nu_i >  =
\sum\limits_{i=1}^{3+n_{s}} {\cal U}_{\alpha i}^\ast | \nu_i > +
\sum\limits_{i=3+n_{s}+1}^{3+n_{s}+n_{R}} {\cal  V}_{\alpha
i}^\ast | \nu_i
>, \label{ch} \ee
\\
we can express the Hamiltonian Eq.~\ref{fhint} in
the flavour representation.

If no sterile neutrinos are present ($n_s=0$), but some heavy neutrinos exist, 
$n=3+n_R$, $n_R \geq 1$, then  ($\alpha, \beta
= e , \mu , \tau $)

\be H_{\alpha \beta}=\left( {\cal U} \frac{\Delta m^2}{2E} {\cal
U}^\dagger \right)_{\alpha \beta }+ \sqrt{2} G_F  \left[ N_e (
\delta_{\alpha e} \delta_{\beta e}- c_{\alpha \beta}
(\delta_{\alpha e} + \delta_{\beta e}) + c_{\alpha e} c_{e
\beta}) +  \frac 12 N_n ( 2 c_{\alpha \beta} - \sum\limits_{\gamma=e,\mu,\tau}
c_{\alpha \gamma} c_{\gamma  \beta} )  \right]. \label{l0} \ee

If a single sterile neutrino ($n_s=1$) and some heavy
neutrinos exist, $n=4+n_R$, $n_R \geq 1$, then ($\alpha, \beta = e , \mu , \tau , s$)

\bea H_{\alpha \beta}& = & \left( {\cal U} \frac{\Delta m^2}{2E}
{\cal U}^\dagger \right)_{\alpha \beta } \label{l1}\\
&+& \sqrt{2} G_F  \left[ N_e \left\{  \delta_{\alpha e}
\delta_{\beta e}- c_{\alpha \beta} (\delta_{\alpha e} +
\delta_{\beta e}) + c_{e \alpha} c_{e \beta} \right\} +  \frac 12
N_n \left\{  \delta_{\alpha s} \delta_{\beta s}+  c_{\alpha
\beta} ( 2- \delta_{\alpha s} -\delta_{\beta s})-
\sum\limits_{\gamma=e,\mu,\tau} c_{\alpha \gamma} c_{\gamma
\beta} \right\} \right]. \nn \eea

In both above cases (Eqs.~\ref{l0},\ref{l1}), we present
the Hamiltonian in the light neutrino subspace only. The full Hamiltonian
is more complicated and contains parts related to the light-heavy neutrino
mixing. In the full basis, it is represented by a $(3 + n_{s}
+ n_{R}) \times (3 + n_{s} + n_{R})$ dimensional matrix

\be
H_{\alpha \beta} \to
\left(
\begin{array}{ll}
 {\cal U} H^{eff} {\cal U}^{\dagger} & {\cal U} H^{eff} {\cal V}^{'\dagger} \\
{\cal V}^{'} H^{eff} {\cal U}^{\dagger} & {\cal V}^{'} H^{eff} {\cal V}^{'\dagger} \\
\end{array}
\right)_{\alpha \beta}.
\label{fh}
\ee

where $H^{eff}$ is the $(3 + n_{s}) \times (3 + n_{s})$ dimensional
Hamiltonian given by Eqs.~\ref{h1},\ref{hint}, and Eq.~\ref{l0} or Eq.~\ref{l1} is the most upper left part of the
matrix Eq.~\ref{fh}. 
Thus, even though we consider the problem of light neutrinos
propagation only, in the flavour basis Eq.~\ref{ch} we have to deal
with $(3 + n_{s} + n_{R}) \times (3 + n_{s} + n_{R})$ dimensional matrices.

It should be stressed here that, up to now, all Hamiltonians
in Eqs.~\ref{h1}-\ref{fh} are represented by hermitian matrices.
The transformations in Eqs.~\ref{ufbase},\ref{ch} are unitary as we sum over all $n$
available neutrino eigenmass states.

If we now take into consideration the fact that, due to kinematical reasons,
no heavy mass eigenstates in Eq.~\ref{ch} can experimentally be produced,
then the properly normalized states $| \tilde{\nu}_\alpha>$, which
correspond to neutrinos produced in real experiments, are

\be | \tilde{\nu}_\alpha> = \lambda_{\alpha}^{-1}
\sum\limits_{i=1}^{3+n_s} {\cal U}^\ast_{\alpha i} | \nu_i > =
\sum\limits_{i=1}^{3+n_s} \tilde{{\cal U}}_{\alpha i}^\ast | \nu_i >,
\label{real}
\ee
\\
where $\lambda_{\alpha}=\sqrt{\sum_{i=1}^{3+n_s} |{\cal U}_{\alpha
i}|^2}=\sqrt{1-c_{\alpha \alpha}}$
and
$\tilde{{\cal U}}_{\alpha i} = \lambda_{\alpha}^{-1}{\cal
U}_{\alpha i}$. Such states are not orthogonal \be <
\tilde{\nu}_\alpha | \tilde{\nu}_\beta > \neq 0, \;\;\; \alpha
\neq \beta . \label{nort} \ee

Let us notice \cite{val,acta} that in this case the usual notion of flavour neutrinos loose its meaning.
For example, a neutrino which is created with an electron, and is described by the state 
$ | \tilde{\nu}_e> $, can produce besides an electron also a muon or a tau. 
It is better then to see 
active neutrinos as particles which are produced together with charged leptons of particular flavours
(in some charged current weak decays), rather than particles having their own flavours.

When we write Eq.~\ref{s1} in the basis of experimentally
accessible states $| \tilde{\nu}_\alpha>$, we get

\be
i \frac{d}{dt} < \tilde{\nu}_\alpha | \Psi (t) > =
\sum\limits_\beta \tilde{H}_{\alpha \beta} <  \tilde{\nu}_\beta | \Psi (t) >.
\label{l2}
\ee

The Hamiltonian $\tilde{H}_{\alpha \beta}$ is given by ($\lambda_{\alpha \beta}=\lambda_\alpha
\delta_{\alpha \beta}$)

\begin{equation}
\tilde{H} = {\tilde{{\cal U}}} H {\tilde{{\cal U}}^{^{-1}}} = \frac{1}{2E} \tilde{{\cal U}} \left(
\begin{array}{ccccc}
0 & 0 & 0 & 0 & \cdots \\ 0  & \delta m_{21}^2 & 0 & 0 & \cdots
\\ 0 & 0 & \delta m_{31}^2 & 0 & \cdots \\
0& 0& 0& \delta m_{41}^2 & \cdots \\
 \cdots & \cdots & \cdots & \cdots & \ddots 
\end{array}
\right) {\tilde{{\cal U}}^{^{-1}}} + \sqrt{2} G_F
\tilde{{\cal U}} \tilde{{\cal U}}^\dagger \lambda^2 \left(
\begin{array}{ccccc}
 (N_e-\frac{N_n}{2})  & 0 & 0 & 0 & \cdots \\ 0  & -\frac{N_n}{2} & 0 & 0 & \cdots \\  0 & 0 & -\frac{N_n}{2} & 0 & \cdots 
\\ 0& 0& 0& 0  & \cdots \\
 \cdots & \cdots & \cdots & \cdots & \ddots 
\end{array}
 \right) .
\end{equation}

This  matrix, in contrast to the previous hermitian representations
(in Eqs.~\ref{h1}-\ref{fh}),  is not hermitian. 
In practice we can solve
the basic Eq.~\ref{s1} either in the experimental $| \tilde{\nu}_\alpha>$
or in the eigenmass $| \nu_i>$ basis (in both cases we deal with
Hamiltonians having dimensions $(3 + n_{s}) \times (3 + n_{s})$).

Let us assume that at
time $t=0$ the state $|\Psi(0)>=|\tilde{\nu}_\alpha (0)>$ is
produced. In order to find the state $| \tilde{\nu}_\alpha (t)>$, we
need to solve Eq.~\ref{l2} with an initial condition

\be < \tilde{\nu}_\beta | \tilde{\nu}_\alpha (0)> =
(\lambda_\alpha  \lambda_\beta)^{-1} ( \delta_{\alpha \beta} - c_{\beta \alpha}) .
\label{a2}
\ee

In the eigenmass basis we have

\be \label{ev-eq} i \frac{d}{dt} < \nu_k | \tilde{\nu}_\alpha(t)
> = \sum\limits_{i=1}^{3+n_s} H_{ki} <  \nu_i |
\tilde{\nu}_\alpha (t) >, \label{herm} \ee

where

\begin{equation}
 H =  \frac{1}{2E} \left(
\begin{array}{ccccc}
0 & 0 & 0 & 0 & \cdots \\ 0  & \delta m_{21}^2 & 0 & 0 & \cdots
\\ 0 & 0 & \delta m_{31}^2 & 0 & \cdots \\
0& 0& 0& \delta m_{41}^2 & \cdots \\
 \cdots & \cdots & \cdots & \cdots & \ddots 
\end{array}
\right)  + \sqrt{2} G_F
{\cal U}^\dagger  \left(
\begin{array}{ccccc}
 (N_e-\frac{N_n}{2})  & 0 & 0 & 0 & \cdots \\ 0  & -\frac{N_n}{2} & 0 & 0 & \cdots \\  0 & 0 & -\frac{N_n}{2} & 0 & \cdots 
\\ 0& 0& 0& 0  & \cdots \\
 \cdots & \cdots & \cdots & \cdots & \ddots 
\end{array}
 \right) {\cal U}.
\end{equation}

and the initial condition is
\be < \nu_k | \tilde{\nu}_\alpha (0)> = \lambda_{\alpha}^{-1}
{\cal U}^\ast_{\alpha k} \, . \label{a1} \ee

For neutrino propagation in a uniform density medium we can
solve the evolution equation analytically. Then it is simpler
to use the eigenmass basis in which the effective Hamiltonian $H$
is hermitian. We will follow this approach in the next chapter. 
In the case of a medium with varying density, 
Eq.~\ref{l2} or Eq.~\ref{ev-eq}  must usually be solved numerically.
Then any of them, hermitian Eq.~\ref{herm} or non-hermitian Eq.~\ref{l2},
with appropriate initial conditions,  from respectively Eq.~\ref{a1} or Eq.~\ref{a2}, can be used.

Finally, we should also remember that, the nonorthogonality of the $| \tilde{\nu}_\alpha >$
states has some impact on theoretically calculated cross sections for neutrino
production and detection.
Let us consider, for example, a charged lepton production process
$\nu_{\alpha} X \to l^{-}_{\beta} Y$. The production amplitude is then

\be
A(\nu_{\alpha} X \to l^{-}_{\beta} Y) =
\lambda_{\alpha}^{-1} \sum\limits_{i=1}^{3+n_s} {\cal U}^{*}_{\alpha i}  A(\nu_{i} X \to
l^{-}_{\beta} Y) \simeq
\lambda_{\alpha}^{-1} \sum\limits_{i=1}^{3+n_s} {\cal U}^{*}_{\alpha i} {\cal U}_{\beta i}  A(\nu_{m=0} X
\to l^{-}_{\beta} Y)
\ee

The $A(\nu_{m=0} X \to l^{-}_{\beta} Y)$ is simply the Standard Model's
amplitude describing the process $\nu_{\beta} X \to l^{-}_{\beta} Y$.
Thus, the production cross section is

\be
\sigma(\nu_{\alpha} X \to l^{-}_{\beta} Y) = \lambda_{\alpha}^{-2} |\delta_{\alpha \beta} - c_{\alpha \beta}|^2 \sigma^{SM}(\nu_{\beta} X \to l^{-}_{\beta} Y)
\ee

where $\sigma^{SM}(\nu_{\beta} X \to l^{-}_{\beta} Y)$ is the
Standard Model's cross section for the process $\nu_{\beta} X \to l^{-}_{\beta} Y$.
For other processes, which we do not describe here (like, for example,
neutrino elastic scattering), even more complicated "scaling" factors appear.

\section{Oscillations of light neutrinos with  CP and T violating effects}

In what follows  oscillations of three light neutrinos
with $n_R \geq 1$ non-decoupling heavy
neutrinos will be considered. If heavy neutrinos decouple
then the mixing between flavour $\nu_\alpha = ( \nu_e , \nu_\mu ,
\nu_\tau )$ and mass $\nu_i = ( \nu_1 , \nu_2 , \nu_3 )$
neutrinos is described by the $3 \times 3$ unitary  matrix $U$,
namely,

\be
| \nu_\alpha > = \sum\limits_{i=1}^{3} U_{\alpha i}^{\ast} | \nu_i >.
\ee

For the matrix $U$  the standard  parametrization is taken

\be
U=
\left(
\begin{array}{ccc}
U_{e1} & U_{e2} & U_{e3} \\
U_{\mu 1} & U_{\mu 2} & U_{\mu 3} \\
U_{\tau 1} & U_{\tau 2} & U_{\tau 3}
\end{array}
\right)
=\left(
\begin{array}{ccc}
c_{12}c_{13} & s_{12}c_{13} & s_{13}e^{-i \delta} \\
-s_{12}c_{23}-c_{12}s_{23}s_{13}e^{i \delta} & c_{12}c_{23}-s_{12}s_{23}s_{13}e^{i\delta} & s_{23}c_{13} \\
s_{12}s_{23}-c_{12}c_{23}s_{13}e^{i \delta} & -c_{12}s_{23}-s_{12}c_{23}s_{13}e^{i\delta} & c_{23}c_{13}
\end{array}
\right) \, .
\label{u}
\ee

Now, in order to implement effects of heavy neutrinos, as discussed in the
Introduction (Eqs.~\ref{U-first},\ref{c-first}), the submatrix
${\cal V}$ must be introduced. As the number  $n_R$ is unknown, ${\cal
V}$ will be parametrized in a simplified way, using a single
effective heavy neutrino state. In this way, the effective light-heavy
neutrino mixing can be characterized \cite{acta} by three real parameters
$\epsilon_e, \epsilon_\mu, \epsilon_\tau$, and two new (real) phases $ \chi_1,
\chi_2$ (responsible for additional  CP and T asymmetry effects)

\be {\cal  V}=\left(
\begin{array}{c}
\epsilon_e \\
e^{-i \chi_1} \epsilon_\mu \\
e^{-i \chi_2} \epsilon_\tau
\end{array}
\right) \, .
\label{eps}
\ee

Then
$\lambda_{\alpha}=\sqrt{1-\epsilon_\alpha^2}\approx1-\epsilon_\alpha^2/2$
and the $3 \times 3$ matrix ${\cal U}(\epsilon_\alpha)$, to the order
of ${\cal{O}} ( \epsilon^2_\alpha )$, has the form

\be {\cal U}= \left(
\begin{array}{ccc}
U_{e1} \lambda_e , & U_{e2} \lambda_e, & U_{e3} \lambda_e  \\
&& \\
U_{\mu 1} \lambda_\mu - U_{e1}e^{-i \chi_1}  \epsilon_e\epsilon_\mu , \;\;\;
& U_{\mu 2}  \lambda_\mu - U_{e2} e^{-i \chi_1}  \epsilon_e\epsilon_\mu  , \;\;\;
& U_{\mu 3} \lambda_\mu- U_{e3} e^{-i \chi_1}
\epsilon_e\epsilon_\mu   \\
&& \\
U_{\tau 1}   \lambda_\tau -U_{e1}e^{-i \chi_1}   \epsilon_e \epsilon_\tau 
& U_{\tau 2} \lambda_\tau -U_{e2}e^{-i \chi_1}  \epsilon_e \epsilon_\tau 
& U_{\tau 3}\lambda_\tau -U_{e3}e^{-i \chi_1}   \epsilon_e \epsilon_\tau  \\
 - U_{\mu 1}e^{i( \chi_1-\chi_2)}  \epsilon_\mu \epsilon_\tau , & -  U_{\mu 2}e^{i( \chi_1-\chi_2)}  
\epsilon_\mu \epsilon_\tau  , &  - U_{\mu 3}e^{i( \chi_1-\chi_2)}  \epsilon_\mu \epsilon_\tau
\end{array}
\right) \, , \label{uszur}
\ee

where $U_{\alpha i}$ are the unitary mixing matrix elements as in  Eq.~\ref{u}.

Experimental data restrict the light-heavy neutrino mixing elements $\epsilon_\alpha^2$ \cite{review,lang}

\be
\epsilon_e^2 < 0.0054, \;\;\; \epsilon_\mu^2 < 0.0096, \;\;\; \epsilon_\tau^2 < 0.016 \, ,
\label{eps1}
\ee
altogether with  their products
\be
\epsilon_e \epsilon_\mu  < 0.0001, \;\;\; \epsilon_\mu \epsilon_\tau < 0.01.
\label{eps2}
\ee

There are no constraints on additional  CP breaking phases $\chi_1$ and $\chi_2$, so we
will assume, as in the case of the standard CP phase $\delta$
\cite{domain}, that any values in the range $(0, 2 \pi \rangle$ are
possible. The matrix $\Omega_{ki}$ in Eq.~\ref{LNC} is given by ($R$
stands for the single effective heavy neutrino)

\be \Omega_{ki}=\delta_{ki}-{\cal  V}_{Rk}'^\ast {\cal  V}_{Ri}',
\;\;\; k,i=1,2,3, \ee and \be {\cal
V}_{Ri}'=-U_{ei}\epsilon_e-U_{\mu i} e^{i \chi_1}
\epsilon_\mu-U_{\tau i} e^{i \chi_2} \epsilon_\tau . \ee

Now, assuming a constant matter density, the evolution equation
Eq.~\ref{ev-eq} can be solved analytically. First, the hermitian
effective Hamiltonian
\be H_{ki}^{eff}=\frac{1}{2E} \left( m_i^2 \delta_{ki} + 2
\sqrt{2} G_F E \left( N_e {\cal U}_{ek}^\ast {\cal U}_{ei} -
\frac{1}{2} N_n \Omega_{ki} \right) \right) \ee
can be diagonalized by a unitary transformation

\be
H^{eff}=\frac{1}{2E} \tilde{W}^\dagger diag ( \tilde{m}_i^2 ) \tilde{W},
\ee
 where  $ \tilde{m}_i^2$ are (real) eigenvalues of $2 E H^{eff}$ and $\tilde{W}$ is  a matrix build of eigenfunctions of $2 E H^{eff}$.
Then Eq.~\ref{ev-eq} takes the form
\be
i \frac{d}{dt} \Psi^\alpha (t) = \frac{1}{2E} \tilde{W}^\dagger diag(\tilde{m}_i^2) \tilde{W} \Psi^\alpha(t),
\ee
where (the $| \tilde{\nu}_\alpha >$ states are given by Eq.~\ref{real})
\be
\Psi^\alpha(t) = \left(
\begin{array}{c}
<\nu_1 | \tilde{\nu}_\alpha(t)> \\
<\nu_2 | \tilde{\nu}_\alpha(t)> \\
<\nu_3 | \tilde{\nu}_\alpha(t)>
\end{array}
\right).
\ee





This equation together with the initial condition Eq.~\ref{a1} can easily be solved giving
\be \Psi^\alpha_k(t)= \sum\limits_i (\tilde{W}^\dagger)_{ki}
e^{-i \frac{\tilde{m}_i^2}{2E} t} (\tilde{W} \tilde{{\cal
U}}^{\dagger} )_{i \alpha} \, , \ee
and the amplitude
$A_{\alpha \to \beta} (L)$ for $\nu_\alpha \to \nu_\beta$ neutrino oscillations
in matter, after traveling a distance L, is given by
\be
A_{\alpha \to \beta} (L)=< \tilde{\nu}_\beta (0) | \tilde{\nu}_\alpha (L=t) > =
\sum\limits_{i=1}^3 {\overline W}_{\beta i} {\overline W}^{\, \ast}_{\alpha i}  e^{-i \frac{\tilde{m}_i^2}{2E} L}.
\ee
The non-unitary neutrino mixing matrix ${\overline W}$ is defined as
\be {\overline W}_{\alpha i} = ( \tilde{{\cal U}} \tilde{W}^\dagger )
_{\alpha i} = \lambda_{\alpha}^{-1} \sum\limits_k {\cal
U}_{\alpha k} \tilde{W}_{ik}^\ast \equiv \lambda_{\alpha}^{-1} W_{\alpha i}. \ee

The final transition probability $P_{\alpha \to \beta} (L) = \left|  A_{\alpha \to \beta} (L) \right|^2$
is the following
\bea P_{\alpha \to \beta} (L) &=& \frac{1}{\lambda_\alpha^2
\lambda_\beta^2} \left\{ \left(  \delta_{\alpha \beta}- \left|
\left( {\cal  V}{\cal  V}^{\dagger} \right)_{\alpha \beta} \right|
\right)^2 - 4 \sum\limits_{i>k} {R}^{ik}_{\alpha \beta}
\sin^2{\Delta_{ik}} + 8 {I}^{12}_{\alpha \beta}
\sin{\Delta_{21}} \sin{\Delta_{31}} \sin{\Delta_{32}} \right. \nonumber \\
&+& \left. 2 \left[  A^{(1)}_{\alpha \beta} \sin{2 \Delta_{31}}+
A^{(2)}_{\alpha \beta} \sin{2 \Delta_{32}} \right] \right\},
\label{prob}
\eea
where
\bea
{R}^{ik}_{\alpha \beta}&=&Re
\left[ W_{\alpha i} W_{\beta k} W_{\alpha k}^{\ast}  W_{\beta i}^{\ast}  \right] , \label{re} \\
{I}^{ik}_{\alpha \beta}&=&Im
\left[ W_{\alpha i} W_{\beta k} W_{\alpha k}^{\ast}  W_{\beta i}^{\ast}   \right] , \label{im} \\
A_{\alpha \beta}^{(i)}(\epsilon_\alpha) &=& Im \left[ W_{\alpha
i}^\ast W_{\beta i} c_{\alpha \beta}^{*} \right] \label{ai}  , \eea
and
\be
\Delta_{ik} = 1.267 \frac{(\tilde{m}_i^2 - \tilde{m}_k^2) [eV^2]  L [km] }{E [GeV] } \, .
\ee

It is interesting to notice that \cite{acta}, while in the case of unitary mixing matrix $(n_R=0)$ we always have 
$\sum_{\beta=e,\mu,\tau (,s)} P_{\alpha \to \beta}=1$, it is no
longer true when $n_R \geq 1$. In this case, as a consequence of
nonorthogonality of the $| \tilde{\nu}_\alpha >$ states, this sum can
be bigger or smaller than 1,
and its value changes with neutrino energy and distance (time).

In agreement with Eq.~\ref{effh}, the transition probability for Dirac antineutrino or
Majorana neutrino with $\lambda=+1$ can be obtained from Eq.~\ref{prob} after the replacement
\be P_{\bar{\alpha} \to \bar{\beta}} (L) = P_{\alpha \to \beta}
(L; {\cal U} \to {\cal U}^\ast, G_F \to -G_F ) \label{antiprob}. \ee


Then conventional CP and T violation probability differences are 
\bea
\Delta P_{\alpha \to \beta}^{CP}   = P_{\alpha \to \beta} - P_{\bar{\alpha} \to \bar{\beta}} &=&
-2 \frac{1}{\lambda_\alpha^2 \lambda_\beta^2} \biggl\{
2 \sum\limits_{i>k} \biggl[ {R}^{ik}_{\alpha \beta} (G_F)   \sin^2{\Delta_{ik}}  (G_F)- {R}^{ik}_{\alpha \beta} (-G_F)   \sin^2{\Delta_{ik}}  (-G_F) \biggr]
 \nonumber \\
&-& 4 \biggl[ {I}^{12}_{\alpha \beta} (G_F)   \sin{\Delta_{21}}  (G_F) \sin{\Delta_{31}}  (G_F) \sin{\Delta_{32}} (G_F) \nonumber \\
&+&
 {I}^{12}_{\alpha \beta} (-G_F)   \sin{\Delta_{21}}  (-G_F) \sin{\Delta_{31}}  (-G_F) \sin{\Delta_{32}} (-G_F) \biggr] \nonumber \\
&-&  \biggl[  A^{(1)}_{\alpha \beta}  (G_F) \sin{2 \Delta_{31}}  (G_F) + A^{(2)}_{\alpha \beta}  (G_F) \sin{2 \Delta_{32}}  (G_F) \nonumber \\
&+&  A^{(1)}_{\alpha \beta}  (-G_F) \sin{2 \Delta_{31}}  (-G_F) + A^{(2)}_{\alpha \beta}  (-G_F) \sin{2 \Delta_{32}}  (-G_F) \biggr] \biggr\}, \label{dcp}
\eea
and
\bea
\Delta P_{\alpha \to \beta}^{T}   = P_{\alpha \to \beta} -  P_{\beta \to  \alpha} = 4  \frac{1}{\lambda_\alpha^2 \lambda_\beta^2}  \left\{
4 {I}^{12}_{\alpha \beta}  \sin{\Delta_{21}} \sin{\Delta_{31}} \sin{\Delta_{32}}+   A^{(1)}_{\alpha \beta} \sin{2 \Delta_{31}}+
A^{(2)}_{\alpha \beta} \sin{2 \Delta_{32}} \right\}.
\label{pt}
\eea

Nonunitarity of the $W$  matrix produces  two types of
effects. At first, all conventional quantities such as ${R}^{ik}_{\alpha
\beta}, {I}^{ik}_{\alpha \beta}$ depend on $\epsilon_\alpha$ and new
CP phases $\chi_i$. At second,  new terms proportional to $
A^{(i)}_{\alpha \beta}$ appear.

The first effect can mainly be seen in numerical analysis. The presence of the additional term is more spectacular and can be analyzed directly. For neutrino oscillations in  vacuum $(G_F=0)$  new
terms do not change the relation between 
$\Delta P_{\alpha \rightarrow \beta}^{CP}$ and $\Delta P_{\alpha \rightarrow \beta}^{T}$, and they are equal

\be
\Delta P_{\alpha \rightarrow \beta}^{CP} (vacuum) =\Delta P_{\alpha \rightarrow \beta}^{T} (vacuum).
\ee

For $\alpha = \beta$, $A_{\alpha \alpha} ^{(i)}=0$ and

\be
\Delta P^{T}_{\alpha \rightarrow \alpha} (matter)=\Delta P^{T}_{\alpha \rightarrow \alpha} (vacuum) = \Delta P^{CP}_{\alpha \rightarrow \alpha} (vacuum) =0.
\ee

In the  normal medium, which is matter and not antimatter,  
$\Delta P^{CP}_{\alpha \rightarrow \alpha} (matter) \neq 0$.
Heavy neutrinos will make this effect  stronger.
For long baseline (LBL) neutrino oscillations 

\be
\Delta_{LBL} \approx \Delta_{31} \simeq \Delta_{32} \simeq {\cal{O}} (1),\;\;\;\; \Delta_{21} \simeq 0 \, ,
\ee
and, in contrary to the unitary oscillations,
\bea
 \Delta P^{T}_{\alpha \rightarrow \beta} (vacuum) =  \Delta P^{CP}_{\alpha \rightarrow \beta} (vacuum ) & \neq & 0, \label{tcp} \\
 \Delta P^{T}_{\alpha \rightarrow \beta} (matter)
 & \simeq & 4 \frac{1}{\lambda_\alpha^2 \lambda_\beta^2} ( A_{\alpha \beta}^{(1)}+ A_{\alpha \beta}^{(1)}) \sin 2 \Delta_{LBL} \nonumber \\
&=& - 4 \frac{1}{\lambda_\alpha^2 \lambda_\beta^2} Im(W_{\alpha 3}^\ast W_{\beta 3} \, c_{\alpha \beta}^{*}) \sin 2 \Delta_{LBL}. \eea

This means that CP and T asymmetries appear even for two flavour neutrino transitions. 
Furthermore, for unitary 3 flavour neutrino oscillations, moduli of all Jarlskog invariants
are equal and, as a consequence, all T asymmetries (equivalent to CP asymmetries in vacuum) are equal 
as well 

\be
  \Delta P^{T}_{e \rightarrow \mu}= \Delta P^{T}_{\mu \rightarrow \tau} = \Delta P^{T}_{\tau \rightarrow e}.
\ee

In addition, if any element of the mixing matrix is small
(vanishes) then the above asymmetries are also small (vanish). For
${\cal  V} \neq 0$ (and as a consequence ${\cal  V'} \neq 0$), there is

\bea
I_{e \mu}^{ik} &=& I_{\mu \tau}^{ik} + Im[W_{\mu i} W_{\mu k}^\ast ({\tilde W} {\cal  V'}^{T}
{\cal  V'}^{*} {\tilde W}^{\dag})_{ik}] =
I_{\tau e}^{ik} - Im[W_{e i} W_{e k}^\ast ({\tilde W} {\cal  V'}^{T} {\cal  V'}^{*} {\tilde W}^{\dag})_{ik}], \nonumber \\
I_{\alpha \beta}^{12} &=& I_{\alpha \beta}^{23} - Im[W_{\alpha 2}^\ast W_{ \beta 2} c_{\alpha \beta}^{*}]=
-I_{\alpha \beta}^{13} + Im[W_{\alpha 1}^\ast W_{\beta 1} c_{\alpha \beta}^{*}] \, . \label{vv} \eea

The above relations and terms proportional to $ A_{\alpha \beta}^{(i)}$ in Eq.~\ref{pt} imply

\be
 \Delta P^{T}_{e \rightarrow \mu} \neq  \Delta P^{T}_{\mu \rightarrow \tau} \neq  \Delta P^{T}_{\tau \rightarrow e}.
\ee

From Eq.~\ref{vv} follows that even when some of the  $U$ matrix elements vanish, 
the asymmetry $ \Delta P^{T}_{\alpha \rightarrow \beta}$ can be nonzero.
For instance, if $W_{e3}=0$ then $I^{i3}_{e  \beta}=0$ ($i=1,2, \; \beta=\mu, \tau $),
but remaining five CP invariants, where  $W_{e3}$ is absent, do not vanish.


\section{Numerical results}

Here we will present some numerical analysis of the standard and nonstandard CP and T 
violating effects for real future neutrino oscillation experiments. 
In order to check the effect of non-decoupling heavy neutrinos in neutrino oscillations, the CP and T
asymmetries for two baselines L=295 km and L=732 km are
calculated. Neutrino energy is allowed to vary,  according to the experimental conditions, in the range
$E =0.1 \div 30$ GeV for $\nu_e \to \nu_\mu$ and $E =1.78 \div 30$ GeV for $\nu_\mu \to \nu_\tau$.
These baselines are planed for several future experiments (e.g. JHF and SJHF in Japan with $E \sim 1$ GeV
\cite{jhf}; ICARUS \cite{ic} and OPERA \cite{op} in Europe, 
$E \sim 20$ GeV; MINOS \cite{min} in USA, $E \sim 10$ GeV; and SuperNuMi \cite{snu} in USA 
with $E \sim3,7,15$ GeV). In all these experiments neutrino
beams from the pion decays will be used, so they are mostly muon neutrino and antineutrino beams.  
Neutrino factories will give in addition electron neutrino and antineutrino beams. 
In Figs.~{2a}-{6c} the probability difference
$\Delta P_{\alpha \to \beta}^{CP}$  for $\nu_e \to \nu_\mu$ and $\nu_\mu \to \nu_\tau$
is shown. For $\nu_\mu \to \nu_\tau$ also standard quantities defined as

\begin{eqnarray}
A^{CP}_{\mu \to \tau} &=& \frac{ \Delta P_{\mu \to \tau}^{CP}}{P(\nu_\mu \to \nu_\tau)+P(\bar{\nu}_\mu \to \bar{\nu}_\tau)},  
\\
A^{T}_{\mu \to \tau} &=& \frac{ \Delta P_{\mu \to \tau}^{T }}{P(\nu_\mu \to \nu_\tau)+P({\nu}_\tau \to {\nu}_\mu)}
\end{eqnarray}

are presented and  matter effects are included 

\bea
A_e\;[eV^2]\;&=& 2 \sqrt{2} G_F N_e E = 7.63 \times 10^{-5} \left[ \frac{\rho}{g/cm^3} \right] \left[ \frac{Y_e}{0.5} \right] \left[ \frac{E}{GeV} \right], \label{ae} \\
A_n\;[eV^2]\;&=& \sqrt{2} G_F N_n E = 7.63 \times 10^{-5} \left[ \frac{\rho}{g/cm^3} \right] \left[ 
1-{Y_e} \right] \left[ \frac{E}{GeV} \right]. \label{an}
\eea

For L=250 km and L=732 km neutrinos pass only the first shell of the
Earth's interior \cite{irina} with a constant density  $\rho=2.6\; g/cm^3$ and $Y_e=0.494$. 
Then $A_e\; [eV^2]=1.96 \times 10^{-4}\;
[E/GeV]$ and  $A_n\; [eV^2]=1.0 \times 10^{-4}\; [E/GeV]$. 
The oscillation parameters are taken from  the best LMA fit values \cite{best}
$\tan^2{\Theta_{23}}=1.4$, $\delta m_{32}^2=3.1 \cdot 10^{-3}\;eV^2$,
  $\tan^2{\Theta_{12}}=0.36$, $\delta m_{21}^2=3.3 \cdot 10^{-5}\;eV^2$ and $ \tan^2{2 \Theta_{13}}=0.005$. 


The parameters $\epsilon_e$, $\epsilon_\mu$ and $\epsilon_\tau$ (Eq.~\ref{eps}) 
are small and satisfy the experimental
constraints given by Eqs.~\ref{eps1},\ref{eps2}.  In agreement with
the above constraints two sets of $\epsilon_\alpha$ parameters will be discussed

\bea
(A) &:& \epsilon_e \sim 0.001, \;\; \epsilon_\mu = \epsilon_\tau=0.1, \label{sa}\\
(B) &:& \epsilon_e =\epsilon_\mu \sim 0.01, \;\; \epsilon_\tau=0.1. \label{sb}
\eea

Figs.~{2a},{2b} present  $\Delta P_{e \to \mu}^{CP}$ (L=250 km)
 for neutrino oscillations  in vacuum (Fig.~{2a}) and matter (Fig.~{2b}), respectively. 
We can see that matter
effects are very weak. The hatched regions in these figures, and in
 all following figures, describe the classical unitary neutrino oscillations, 
with $0 < \delta \leq 2 \pi$, and they are similar in both cases.
In Fig.~{2b} this region is only slightly asymmetric. As parameters $\epsilon_\alpha$
which describe non-unitary oscillations 
are very small, deviation of  $\Delta P_{e \to \mu}^{CP}$  from unitary oscillations 
is very weak. The shaded regions in Figs.~{2a},{2b}, and in all following figures, give the allowed range of  
$\Delta P_{e \to \mu}^{CP}$  for the non-unitary case with additional
 CP breaking phases $\chi_1,\chi_2$ (Eq.~\ref{eps}) that change in
 the domain $0 < \chi_i \leq 2 \pi$. 
It is interesting to see the same CP violating quantity  $\Delta P_{e \to \mu}^{CP}$ for 
$\Theta_{13} \to 0$. In vacuum this
quantity equals to zero. In Fig.~{2c} a possible range of $\Delta P_{e \to \mu}^{CP}$ is depicted for non-unitary 
oscillations. In agreement with our previous discussion (Eq.~\ref{tcp}) such a quantity does not vanish. However, it is small, which comes out of  strong experimental bounds Eqs.~\ref{sa},\ref{sb}. 
Situation does not change qualitatively 
in the matter case (Fig.~{2d}), except that this time unitary oscillations  can be nonzero. 
Such miserable effects cannot be detected in future neutrino experiments.
Of course, $A^{CP}_{e \to \mu}$ and $A^T_{e \to \mu}$ effects  alone can be large
(see e.g. \cite{acta},\cite{xing}). We would like to stress, however, that  $A^{CP}_{\alpha \to \beta}$ and $A^T_{\alpha \to \beta}$ must be discussed together 
with  $\Delta P_{\alpha \to \beta}^{CP}$ quantities. Only then we can say if these effects can really be measured in experiments. 
Such a discussion 
for $\nu_\mu \to \nu_\tau$ transitions will follow in the next figures.

In Figs.~{3a,3b}  the $\Delta P_{\mu \to \tau}^{CP}$ and
 $A_{\mu \to \tau}^{CP}$  quantities in vacuum  are presented 
as functions of neutrino energy for 
L=250 km and the first set of $\epsilon_\alpha$ parameters from
 Eq.~\ref{sa}. Here  the CP violating effects are quite large. 
For example, for E=2 GeV,  $\Delta P_{\mu \to \tau}^{CP} \in ( -0.0003, +0.0003) $ in the unitary case (hatched region in Fig.~{3a}) and $\Delta P_{\mu \to \tau}^{CP} \in ( -0.017, +0.017) $ in the non-unitary case (shaded region in Fig.~{3a}).
We do not present results for e $A_{\mu \to \tau}^{T}$ here as in vacuum
 $A^T_{\alpha \to \beta}=A^{CP}_{\alpha \to \beta}$.
We can see that for higher neutrino energies non-unitary effects can be very large (and increase 
with neutrino energy). Unfortunately, oscillation probabilities  themselves are getting smaller. 

In Figs.~{4a,4b} the same quantities as in Figs.~{3a,3b} are presented but for the second set of 
$\epsilon_\alpha$ parameters (Eq.~\ref{sb}). As $\epsilon_\mu$ is 10 times smaller, also 
nonunitary effects are smaller by the same factor. 

For L=250 km the effects of neutrino interactions in  the Earth's matter are small
and all presented quantities are almost the same as in vacuum. Thus we do not present them here. 
However, for longer baseline experiments
(L=732 km)  matter effects can already be seen. In  Figs.~{5a,5b} and  Figs.~{6a-c}
 the   $\Delta P_{\mu \to \tau}^{CP}$,
 $A_{\mu \to \tau}^{CP}$ and  $A_{\mu \to \tau}^{T}$ quantities  are presented for $\nu_\mu \to \nu_\tau$ transitions
in vacuum and matter, respectively. The $\epsilon_\alpha$ parameters are taken according to Eq.~\ref{sa}. 
Again, the non-unitary effects are large. 
For the unitary case, the symmetric region of  $\Delta P_{\mu \to
\tau}^{CP}$ in vacuum (Fig.~{5a}) becomes asymmetric (negative) in 
matter (Fig.~{6a}). The full range of  $\Delta P_{\mu \to \tau}^{CP}$ for non-unitary oscillations shifts also slightly 
toward negative values (compare the shaded regions in Fig.~{5a} and  Fig.~{6a}). Once again 
$A_{\mu \to \tau}^{CP}$  (Fig.~{5b} and  Fig.~{6b}) and  $A_{\mu \to \tau}^{T}$ 
 (Fig.~{6c}) asymmetries are very similar to each other. They are larger for higher neutrino energies but, 
unfortunately, not because  $\Delta P_{\mu \to \tau}^{CP}$ is larger, but because  
the probabilities for (anti)neutrino oscillations are getting smaller.

\section{Conclusions}

The evolution equation for neutrinos propagating in matter is found in
the case when heavy neutrinos do not decouple and, as a result, the
mixing matrix between light neutrinos is not unitary.

The neutrino propagation is described by a hermitian Hamiltonian found
in the eigenmass basis of light neutrinos. This basis is the most
convenient to calculate physical effects. Two other basis which we
consider, that means the full orthogonal 
basis which contains both light and heavy neutrino states, and the "experimental"
basis with nonorthogonal light neutrinos only,  are more complicated when applied
to numerics. 
In the first case, the
Hamiltonian is hermitian but has full $(3 + n_{s} + n_{R}) \times (3 +
n_{s} + n_{R})$ dimensions. In the second case, it has smaller $(3 + n_{s})
\times (3 + n_{s})$ dimensions, but is represented by a non-hermitian
matrix. 

If heavy neutrinos do not decouple, the notion of neutrino flavour
loose its meaning. In such case it is better to see active
light neutrinos as particles produced together with charged leptons of particular
flavours rather then particles having their own flavours.
Moreover, the nonorthogonality of such neutrino states has also some impact on theoretically calculated
cross sections for neutrino production and detection.
 
We have found the equation which describes the propagation of light
neutrinos in any polarized, nonhomogeneous, charged matter. Numerical
analysis is done in the simpler case of unpolarized, isotropic, and
electrically neutral matter.

The non-unitary neutrino mixing is especially important for CP and T
violating effects in neutrino oscillation. Two additional CP phases,
which appear in the light-heavy neutrino mixing matrix, have crucial
consequences in such phenomena.

In comparison with normal unitary neutrino oscillation, the CP and T
violation are generally larger and appear even where standard effects
are very small or vanish. The new CP and T violation effects depend on
the strength of light-heavy neutrino mixing. For present experimental
bounds on the non-decoupling parameters, the $\nu_\mu \to \nu_\tau$
oscillation is especially sensitive to the new CP effects. For $\nu_e
\to \nu_\mu$ oscillation, where the bound on the $c_{e \mu}$ parameters
coming from the non observation of the $\mu \to e \gamma$ decay are very
stringent, the effects are smaller. 

In the future, the observation of CP effects (or lack of them) in the
$\nu_\mu \to \nu_\tau$ oscillation can give new information on the
light-heavy neutrino mixing. The bounds on such mixing, found in future
neutrino experiments, can be stronger then the same bounds coming
from charged lepton violating processes.


\section{Acknowledgments}
The work  was supported by the Polish Committee for Scientific Research under 
Grants 2P03B05418 and 2P03B13622.

\vspace{5 cm}

\begin{figure}[ht]
\epsfig{figure=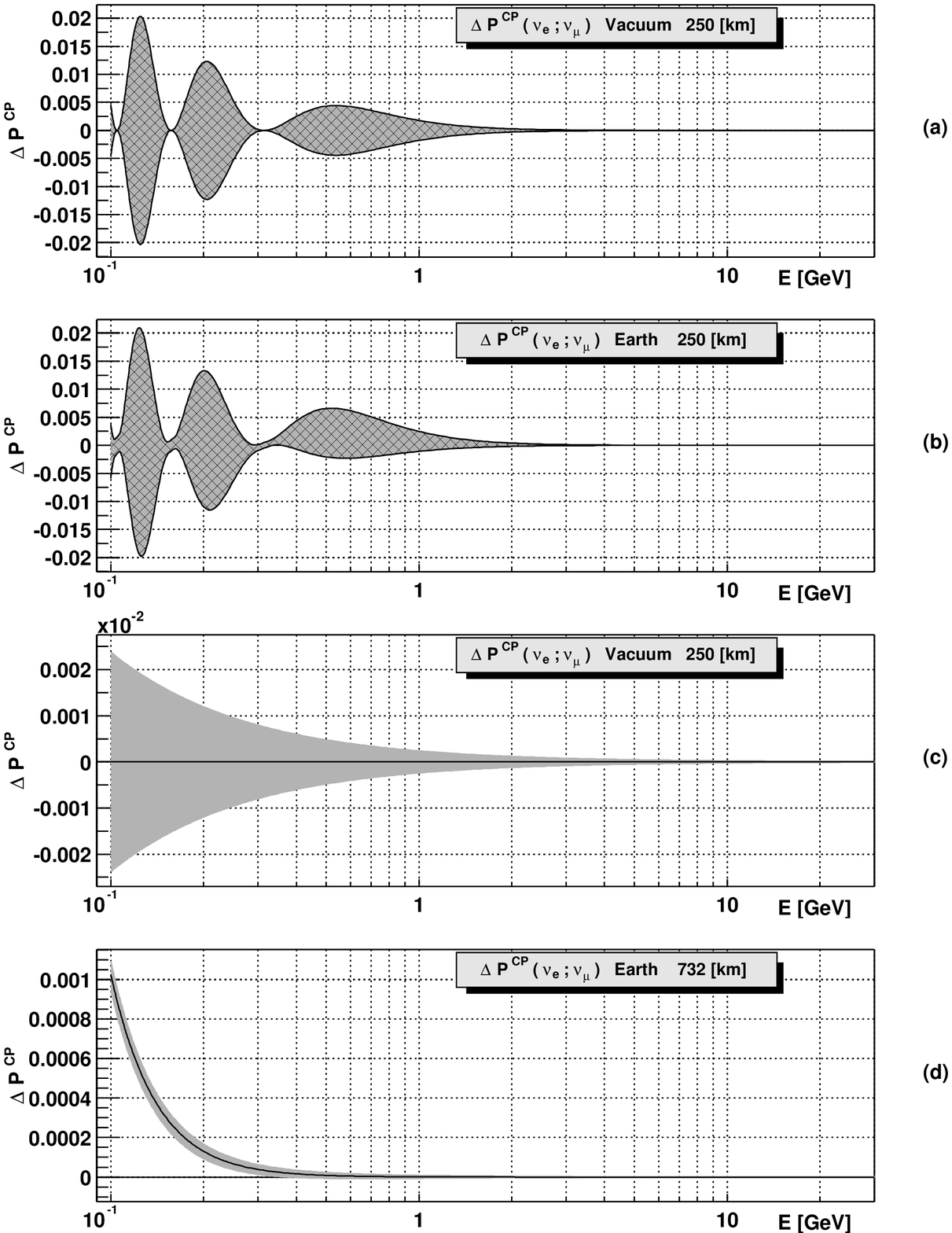, width=13.5cm, height=18cm ,angle=0}
\caption{ \newline
{\bf (a)} The probability difference $\Delta P_{e \to \mu}^{CP}$ defined in Eq.~\ref{dcp},
for L=250 km as a function of neutrino  energy  (oscillations in vacuum). 
The hatched region describes normal unitary oscillations with $0 < \delta \leq 2 \pi$. 
The shaded region corresponds to the non-unitary neutrino oscillations with
additional CP breaking phases $\chi_1,\chi_2$ (Eq.~\ref{eps}) that change in the domain $0 < \chi_i \leq 2 \pi$. The $\epsilon_\alpha$
parameters for both sets from Eqs.~\ref{sa},\ref{sb} give $\epsilon_e \epsilon_\mu =10^{-4}$ and the
results are the same for both cases. Other oscillation parameters  correspond to the best LMA fit values given in \cite{best}
(see the text for details). 
 \newline
{\bf (b)} Neutrino oscillations in matter with the same parameters as in 
Fig.~{2a}. In this case neutrinos pass only the first shell of the
Earth's interior \cite{irina} with a constant  density $\rho = 2.6\;g/cm^3$
and $Y_e=0.494$.
 \newline
{\bf (c)} Neutrino oscillations in vacuum with the same parameters as in 
Fig.~{2a} but for $\Theta_{13}=0$. In this case, in vacuum 
$\Delta P_{\alpha \to \beta}^{CP}=0$
for normal unitary neutrino oscillations. Non-zero effects are exclusively related to the non-decoupling heavy neutrinos. \newline
{\bf (d)} $\Delta P_{e \to \mu}^{CP}$ for neutrino oscillations in matter for $\Theta_{13}=0$, L=732 km.
Other parameters as in Fig.~{2a}. Because of matter effects, 
$\Delta P_{e \to \mu}^{CP}  \neq 0$ even for unitary neutrino oscillations.
\label{1a}}
\end{figure}

\begin{figure}[ht]
\epsfig{figure=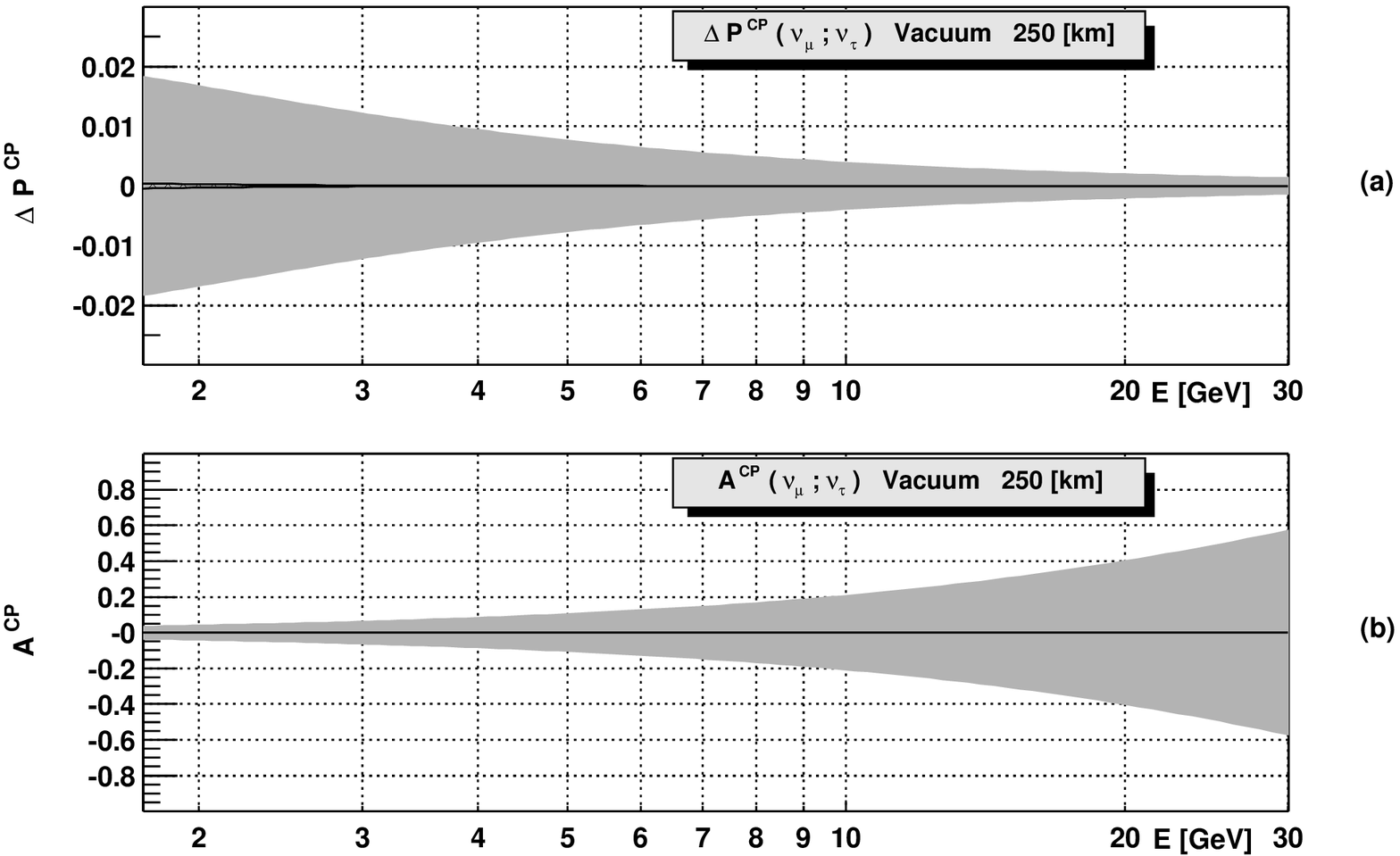, width=13.5cm,  height=9cm, angle=0}
\caption{ 
$\Delta P_{\mu \to \tau}^{CP}$ {\bf (a)} and
$A_{\mu \to \tau}^{CP}$ {\bf (b)}
for neutrino oscillations in vacuum, L=250 km.
Other parameters as in Fig.~{2a}, $\epsilon_\alpha$  according to Eq.~\ref{sa}. 
\label{2a}}
\end{figure}

\begin{figure}[ht]
\epsfig{figure=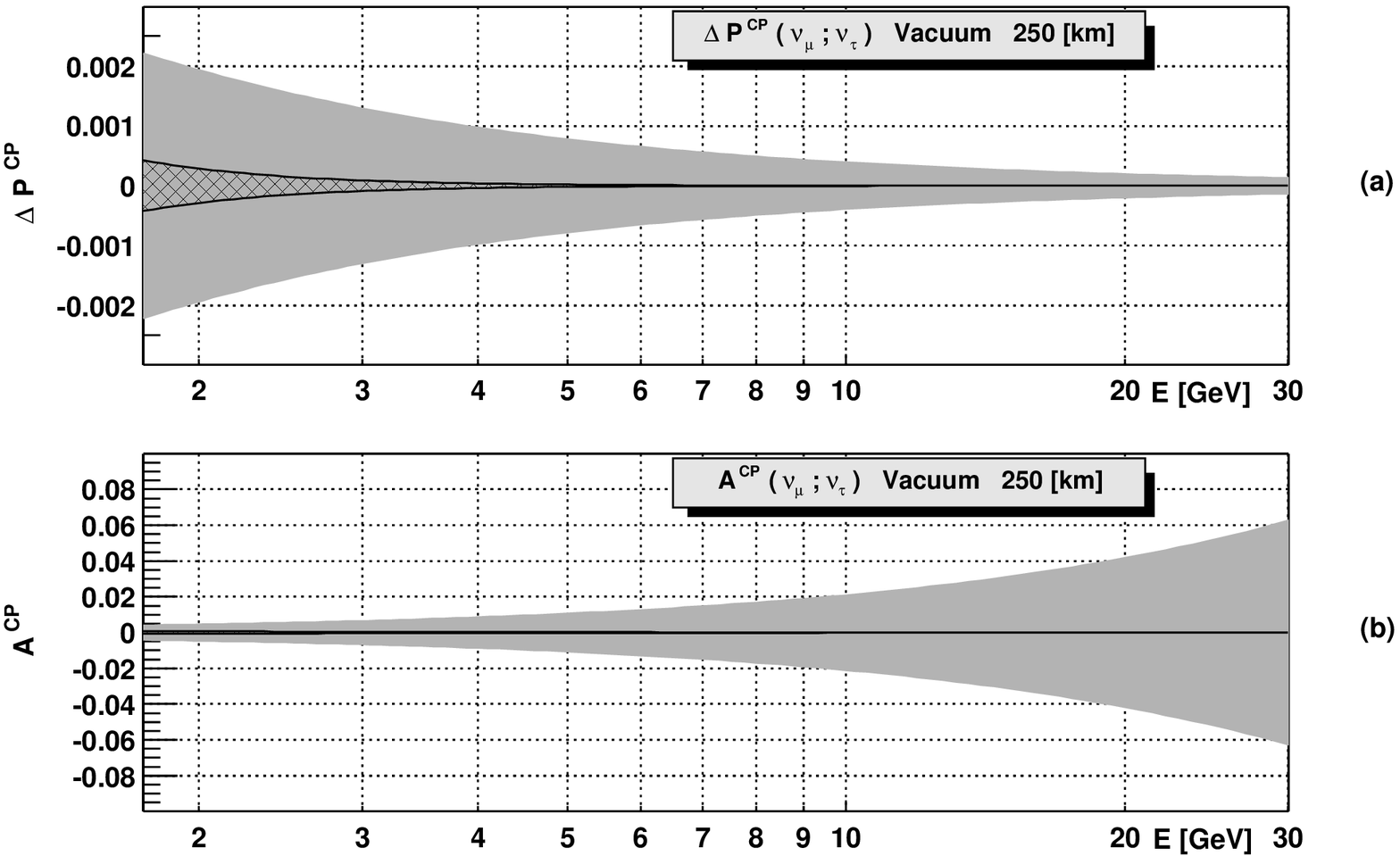, width=13.5cm,  height=9cm, angle=0}
\caption{ 
$\Delta P_{\mu \to \tau}^{CP}$ {\bf (a)} and
$A_{\mu \to \tau}^{CP}$ {\bf (b)} for neutrino oscillations in vacuum, L=250 km.
Other parameters as in Fig.~{2a}, $\epsilon_\alpha$  according to Eq.~\ref{sb}. 
\label{3a}}
\end{figure}

\begin{figure}[ht]
\epsfig{figure=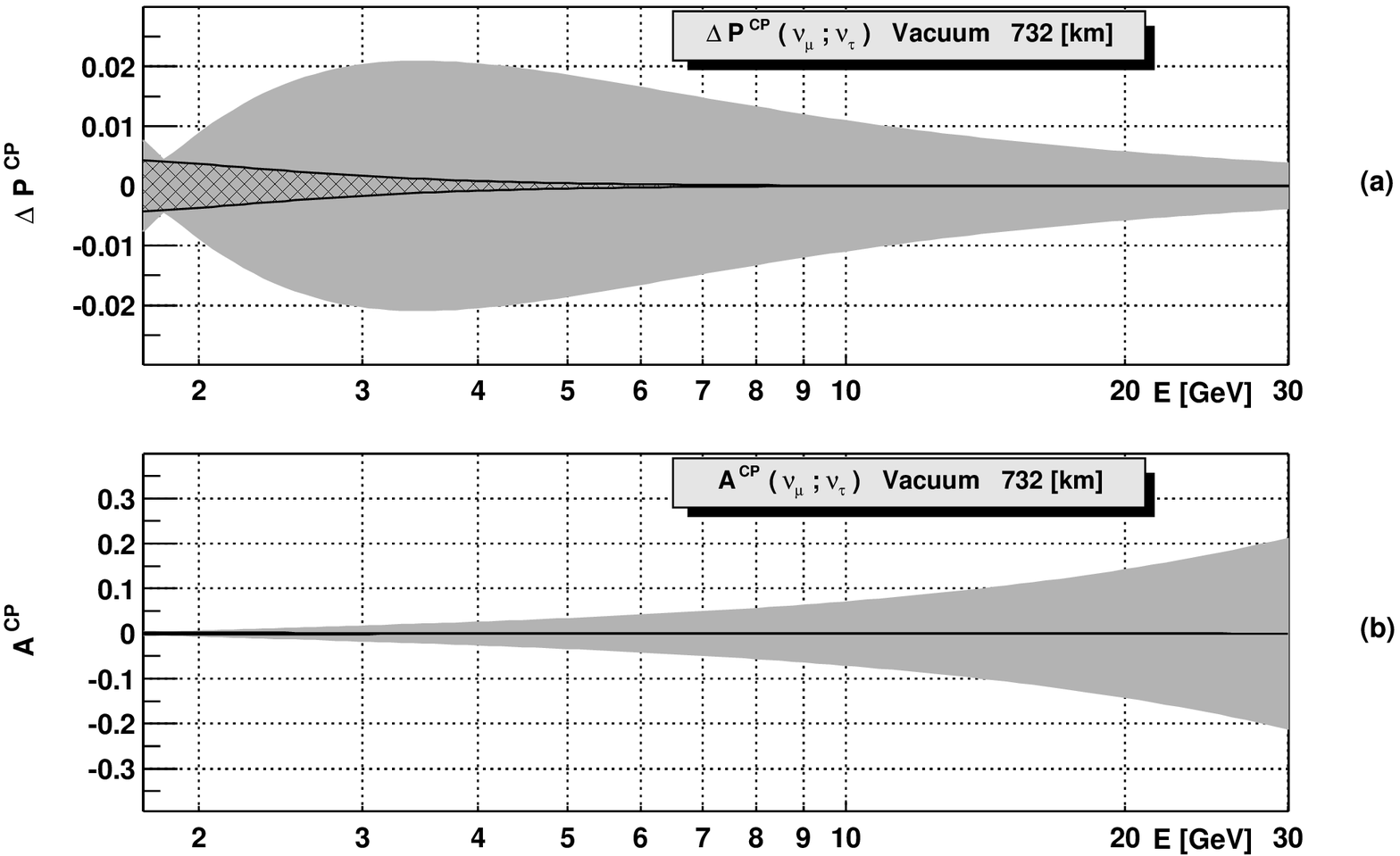, width=13.5cm,  height=9cm, angle=0}
\caption{ 
$\Delta P_{\mu \to \tau}^{CP}$ {\bf (a)} and
$A_{\mu \to \tau}^{CP}$ {\bf (b)}
for neutrino oscillations in vacuum, L=732 km.
Other parameters as in Fig.~{2a}, $\epsilon_\alpha$  according to Eq.~\ref{sa}. 
\label{4a}}
\end{figure}

\begin{figure}[ht]
\epsfig{figure=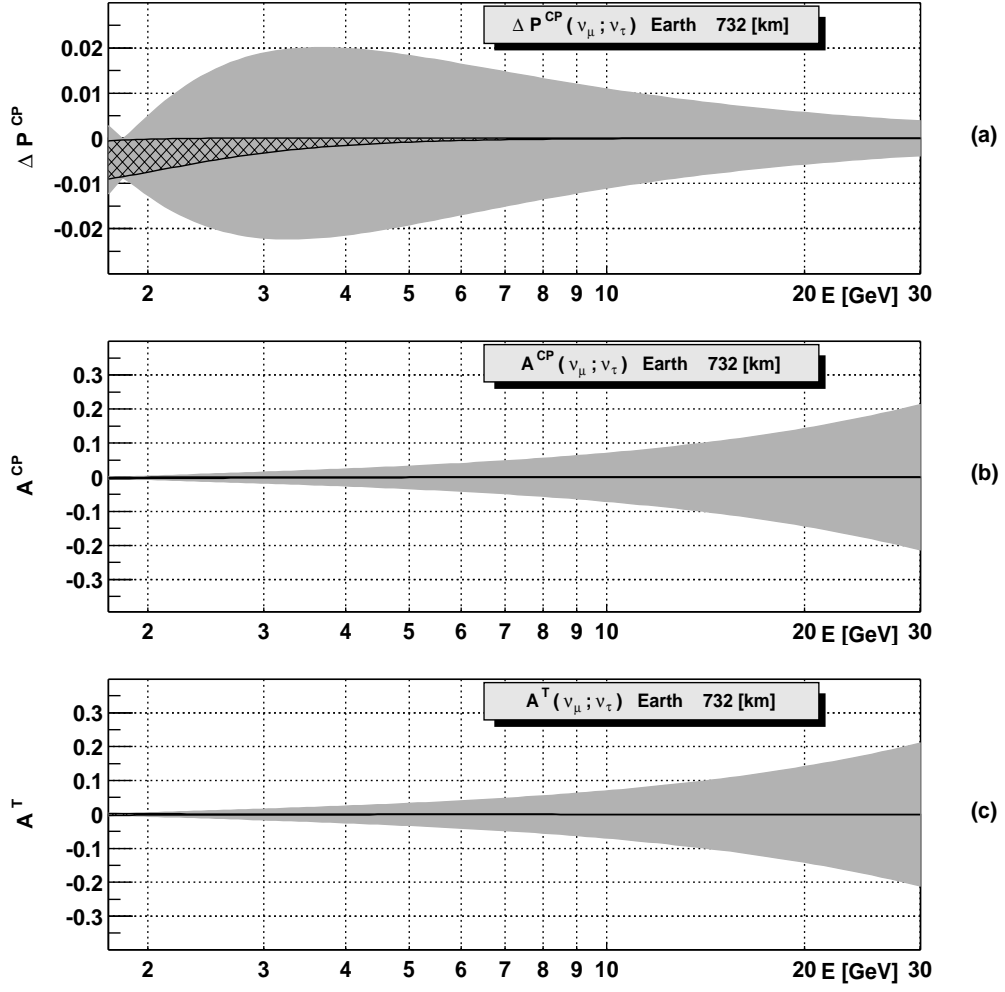, width=13.5cm,  height=13.5cm, angle=0}
\caption{ 
$\Delta P_{\mu \to \tau}^{CP}$ {\bf (a)},
$A_{\mu \to \tau}^{CP}$ {\bf  (b)}, and
$A_{\mu \to \tau}^{T}$ {\bf (c)}
for neutrino oscillations in matter, L=732 km.
Other parameters as in Fig.~{2a}, $\epsilon_\alpha$  according to Eq.~\ref{sa}. 
\label{5a}}
\end{figure}

\end{document}